\documentclass[ floatfix, reprint, superscriptaddress, amsmath,amssymb, aps ]{revtex4-2}

\usepackage{csquotes}
\usepackage{comment}
\usepackage{subfigure}
\usepackage{xcolor}
\usepackage{graphicx}
\usepackage{dcolumn}
\usepackage{bm}
\usepackage{hyperref}
\usepackage{wasysym}

\begin{document}
\title{Frequency-offset separated oscillatory fields technique applied to neutrons}





\author{Anastasio~Fratangelo}
\email[Corresponding author: ]{anastasio.fratangelo@unibe.ch}
\affiliation{Laboratory for High Energy Physics and Albert Einstein Center for Fundamental Physics, University of Bern, 3012 Bern, Switzerland}

\author{Philipp~Heil}
\affiliation{Laboratory for High Energy Physics and Albert Einstein Center for Fundamental Physics, University of Bern, 3012 Bern, Switzerland}

\author{Christine~Klauser}
\affiliation{Paul Scherrer Institut, 5232 Villigen PSI, Switzerland}

\author{Gjon~Markaj}
\affiliation{Laboratory for High Energy Physics and Albert Einstein Center for Fundamental Physics, University of Bern, 3012 Bern, Switzerland}

\author{Marc~Persoz}
\affiliation{Laboratory for High Energy Physics and Albert Einstein Center for Fundamental Physics, University of Bern, 3012 Bern, Switzerland}

\author{Ciro~Pistillo}
\affiliation{Laboratory for High Energy Physics and Albert Einstein Center for Fundamental Physics, University of Bern, 3012 Bern, Switzerland}

\author{Ivo~Schulthess}
\altaffiliation[Present address: ]{Deutsches Elektronen-Synchrotron DESY, 22607 Hamburg, Germany}
\affiliation{Laboratory for High Energy Physics and Albert Einstein Center for Fundamental Physics, University of Bern, 3012 Bern, Switzerland}

\author{Jacob~Thorne}
\affiliation{Laboratory for High Energy Physics and Albert Einstein Center for Fundamental Physics, University of Bern, 3012 Bern, Switzerland}

\author{Florian~M.~Piegsa}
\email[Corresponding author: ]{florian.piegsa@unibe.ch}
\affiliation{Laboratory for High Energy Physics and Albert Einstein Center for Fundamental Physics, University of Bern, 3012 Bern, Switzerland}


\date{\today}
\begin{abstract}
The novel technique of frequency-offset separated oscillatory fields (FOSOF) has been originally proposed as a modification to Ramsey's method of separated oscillatory fields. It has recently been employed in precision measurements with atomic beams since it allows for an alternative approach to determine absolute resonance frequencies.
We present results from a systematic investigation of the FOSOF technique adapted to a beam of cold neutrons. 
\end{abstract}

\maketitle

\section{\label{sec:intro}Introduction}

In the field of cold and ultra-cold neutron physics, Ramsey's method of separated oscillatory fields is widely used \cite{ramsey1949,ramsey1950,RAMSEY1986223}. For instance, the applications encompass the search for the neutron electric dipole moment \cite{PurcellRamsey1950,PhysRevD.15.9,nEDM_2020,beamedm_2013}, the search for exotic interactions and axions \cite{Piegsa2012Axion,Abel2017PRX,Ivo2022}, the measurement of incoherent scattering lengths \cite{Roubeau1974,Abragam1975,Glattli1979,MALINOVSKI1981103,Piegsa2008b}, polarized neutron radiography \cite{Piegsa2009Imaging,PIEGSA20112409}, the measurement of the neutron magnetic moment \cite{gamma_n,PhysRevA.18.1057}, and gravity resonance spectroscopy \cite{PhysRevD.81.065019,QbounceGravity,QbounceGravity1}.
\newline Recently, a modification of Ramsey's original method has been proposed \cite{fosof}. The so-called frequency-offset separated oscillatory fields (FOSOF) technique offers an alternative straight-forward approach to determine absolute resonance frequencies.
It has been successfully applied in a precise measurement of the Lamb shift of atomic hydrogen to determine the proton charge radius and in a measurement of the atomic fine structure of helium \cite{fosof_lamb,fosof_helium}. Here, we demonstrate the realization and systematic characterization of the technique applied to a monochromatic cold neutron beam.

\section{Ramsey and FOSOF technique}


In Ramsey's original method, two oscillating magnetic fields are applied to flip the spins of polarized particles moving in a static homogeneous magnetic field $B_0$. By applying the first oscillating magnetic field pulse, the spins are flipped by $\pi/2$ into the plane perpendicular to the $B_0$ direction where they will start precessing. A second phase-locked oscillating magnetic field pulse is applied after a certain period of free precession time causing a second spin-flip. The two magnetic fields oscillate at the same frequency which is scanned close to the expected resonance. Depending on the applied oscillating frequency, the spins result in the $up$ state, $down$ state, or in a superposition of the two.
Hence, an interference pattern of the spin polarization in the frequency domain is obtained. 
\newline The FOSOF technique differs from Ramsey's method by having the frequencies of the two applied oscillatory fields slightly detuned from each other such that the relative phase of the two fields varies with time. This results in a time-modulation of the detected signal which oscillates with the frequency difference of the two oscillatory fields. 
\newline A FOSOF measurement can be divided into two parts:
an initial scan of the frequency $f$ is performed with the first magnetic field oscillating at $f_1 = f + \delta f/2$, while the second magnetic field oscillates at $f_2 = f-\delta f/2$, where $\delta f$ represents the detuning frequency which is much smaller than $f$. 
In a second scan the frequencies of the oscillating fields are inverted, i.e.\  $f_1 = f - \delta f/2$ and $f_2 = f+\delta f/2$.
Alternatively, this can be interpreted as a change in sign of the detuning frequency. For each frequency $f$, a so-called FOSOF signal is acquired by measuring the particle spin states as a function of the detection time in a time-of-flight type manner. From a sinusoidal fit of such a signal, a corresponding value for the FOSOF phase is extracted. 
Finally, from the phase difference between these two scans, the resonance frequency is determined: its value corresponds to the frequency at which the phase difference becomes zero.

\section{Experimental apparatus}

\begin{figure*}
    \centering
    \includegraphics[scale=0.45]{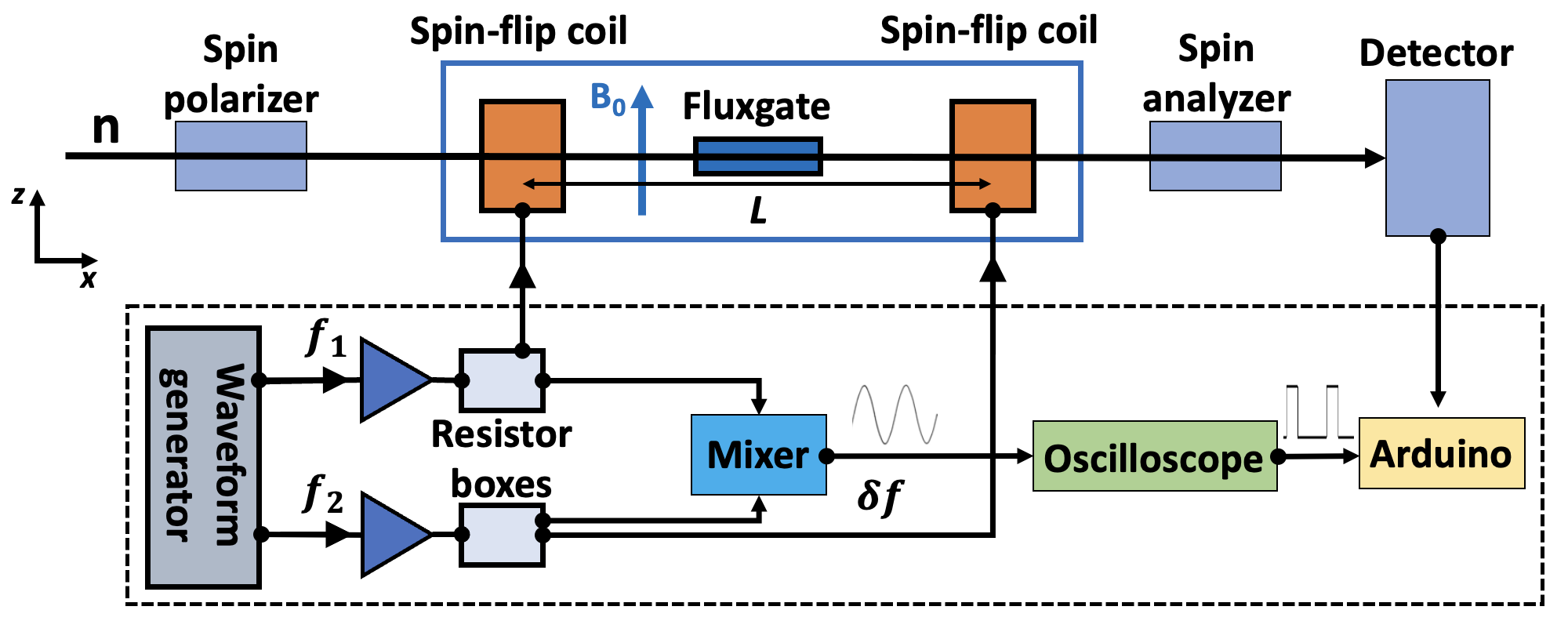}
    \caption{Schematic of the experiment with its main components. 
    The neutron beam (n) oriented along the $x$-direction is spin-polarized and then passes through the FOSOF setup consisting of two spin-flip coils separated by a center-to-center distance $L$ (precession region length). The spin-flip coils produce a linear oscillating field along the beam direction. The vertical magnetic field $B_0$ is monitored by means of a fluxgate magnetometer situated next to the beam path. Finally, the spin states of the neutrons get analyzed before they are counted in a $^3$He-gas detector.
    The electronics and the DAQ system are highlighted by a surrounding dashed rectangle. The core of the DAQ is the Arduino Due micro-controller which collects and counts all the signals in a time-of-flight type measurement.}
    \label{fig:exp_scheme}
\end{figure*}
\begin{figure}
    \centering
    \includegraphics[scale=0.067]{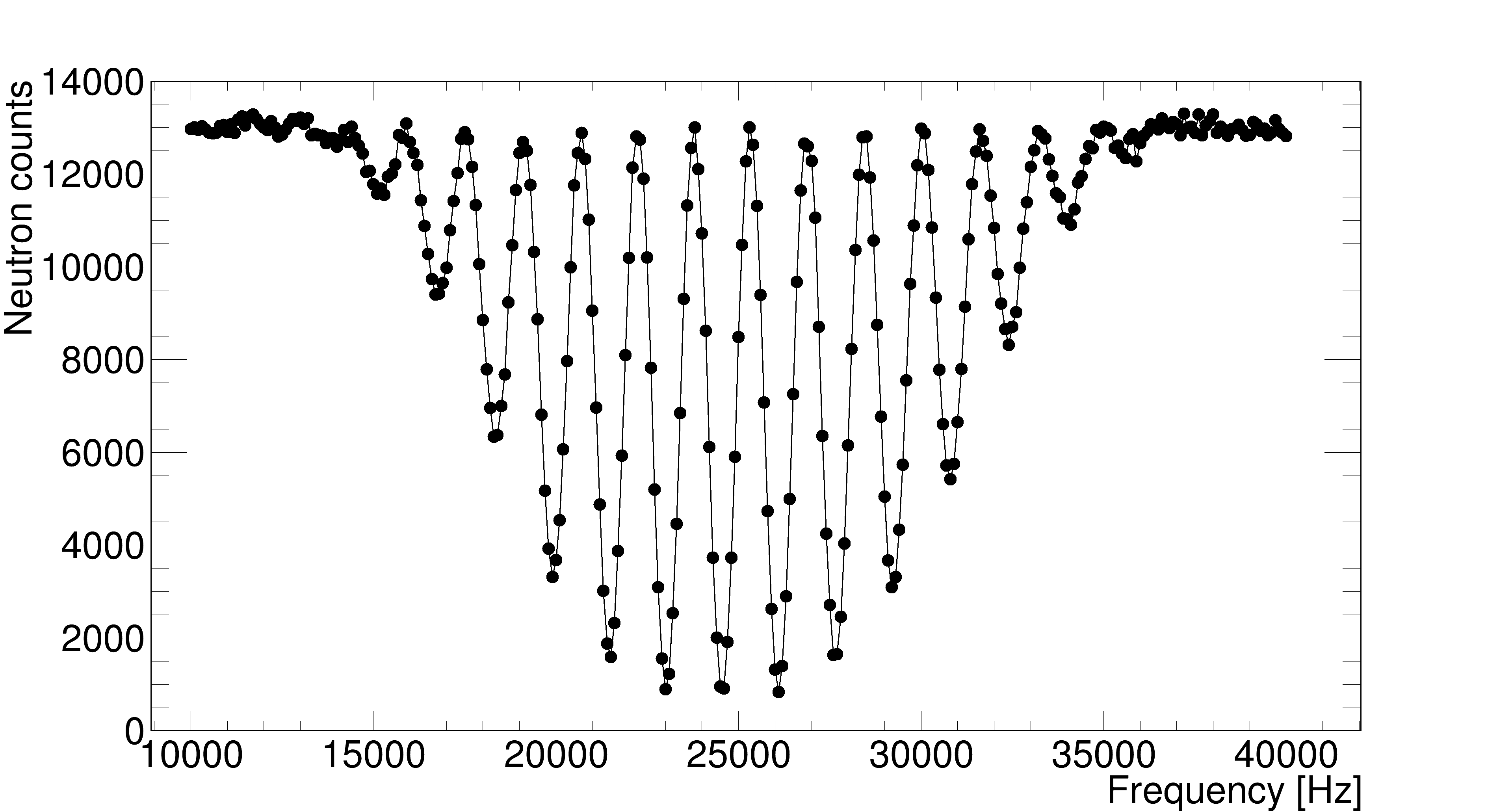}
    \caption{Neutron counts as a function of the spin-flip signal frequency. The data have been obtained with the traditional Ramsey method of separated oscillatory fields. The acquisition time per data point was approximately 5~s. The solid line serves only as a guide for the eyes to highlight the typical Ramsey pattern. 
    }
    \label{fig:ramsey}
\end{figure}

The FOSOF technique has been investigated at the NARZISS neutron reflectometer of the spallation source SINQ at the Paul Scherrer Institut.
The instrument provides a continuous monochromatic cold neutron beam with a de Broglie wavelength of $4.96$ $\text{\AA}$. 
The beamline is equipped with a beam monitor detector which is used to normalize intensities of different measurements compensating  for fluctuations due to potential beam instabilities. 
A schematic of the main components of the experimental apparatus is presented in Fig.\ \ref{fig:exp_scheme}. A set of apertures is used to shape the beam to a width of $2$~mm and a height of $40$~mm. 
The neutrons are polarized with a spin-polarizing supermirror in the direction of the magnetic field $B_0$ (blue arrow pointing in $z$-direction in the figure). 
$B_0$ is generated with the electromagnet of the beamline supplied with a current from a low noise power source (Keysight B2962A). To monitor and actively stabilize the magnetic field, a three-axis fluxgate magnetometer (Sensys FGM3D/1000) is placed next to the neutron beam in the center of the apparatus. A PID control algorithm operates at a 2~Hz rate and stabilizes the magnetic field at the fluxgate position within 18~nT$_{\text{pp}}$ at a nominal set value of $850$~$\mu$T. Note, the fluxgate measures only at one single point in space close to the beam and the stabilization does not account for the entire neutron flight path. For this reason, the magnetic field experienced by the neutrons is slightly different from the stabilized field value (see Sec.\ \ref{sec:mgscan}).
The two spin-flip coils are $80$~mm long and are made from copper wire with a diameter of $0.8$~mm wound around a hollow rectangular shaped support structure made from POM with a cross-section of $30 \times 70$~mm$^2$ (width $\times$ height). Two $2$~mm thick aluminum plates are mounted on either end of the support structure to minimize the magnetic fringe fields. The current for each  spin-flip coil is provided by a waveform generator (Keysight 33600A, with a nominal frequency accuracy of $\pm 1$~ppm) and a $1000$~W audio amplifier (Stageline STA-1000). The amplifiers are connected to the spin-flip coils via a resistor box, containing a $200$~$\Omega$ high-power resistor, to establish a flat frequency response. The resistor boxes provide auxiliary monitor outputs which are connected to a frequency mixer (Mini-Circuits ZAD-8+). Using a low-pass filter on the output of the mixer, one obtains the sinusoidal reference signal of the FOSOF technique which oscillates with the detuning frequency $\delta f$. 
This reference signal is fed into a high-resolution digital oscilloscope (Picoscope 5444B) which generates a pulse every time the signal reaches a predefined threshold. These pulses are employed as trigger signals for the time-of-flight type measurements. 
A spin analyzer placed in front of the detector filters the final spin state of the neutron beam. The neutrons are detected with a $^3$He-gas detector which generates a logic pulse for each detected neutron.  Finally, an Arduino Due micro-controller collects the detector pulses and allocates them in $100$~$\mu$s time bins with respect to the aforementioned trigger signals.
\newline A first test of the apparatus has been performed by conducting a measurement using the traditional Ramsey method of separated oscillatory fields. In this case, the two spin-flip coil signals are phase-locked and oscillate at the same frequency. The neutron counts are plotted as a function of the frequency in Fig.\ \ref{fig:ramsey} describing a typical Ramsey pattern.
The spin-flip signal frequency has been scanned between $10000$~Hz and $40000$~Hz in steps of $100$~Hz \footnote[1]{The audio amplifier has a nominal operational frequency range of up to $20$~kHz, however, it was demonstrated to show good performance up to $45$~kHz.}. The pattern has been measured with a nominal magnetic field at the fluxgate position $B_{fluxgate} = 850$~$\mu$T and a precession region length $L=500$~mm, i.e.\ center-to-center distance between the spin-flip coils. The pattern exhibits a period of about 1600~Hz. Each data point has been measured with $10^5$ neutron counts in the beam monitor detector corresponding to approximately 5~s of acquisition time. 
From considerations regarding the applied magnetic field and in accordance with independently performed Rabi measurements, we determined the central minimum at about $24500$~Hz to be the actual Larmor resonance frequency.
Hence, with the FOSOF technique, we expect to identify the resonance close to this frequency value.

\section{FOSOF oscillating signal}
\begin{figure} 
    \subfigure[]{\includegraphics[scale=0.17]{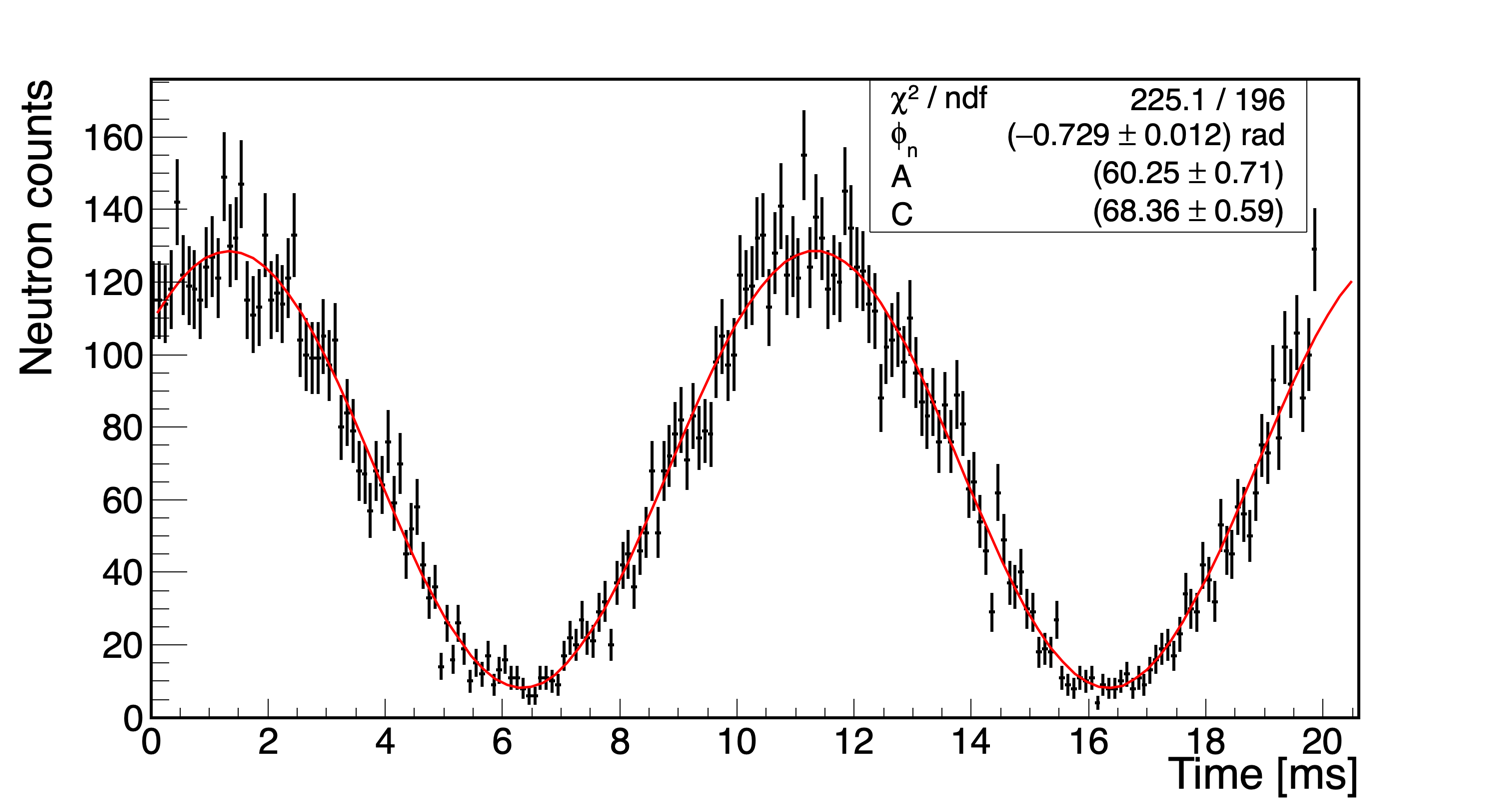}}
    \subfigure[]{\includegraphics[scale=0.17]{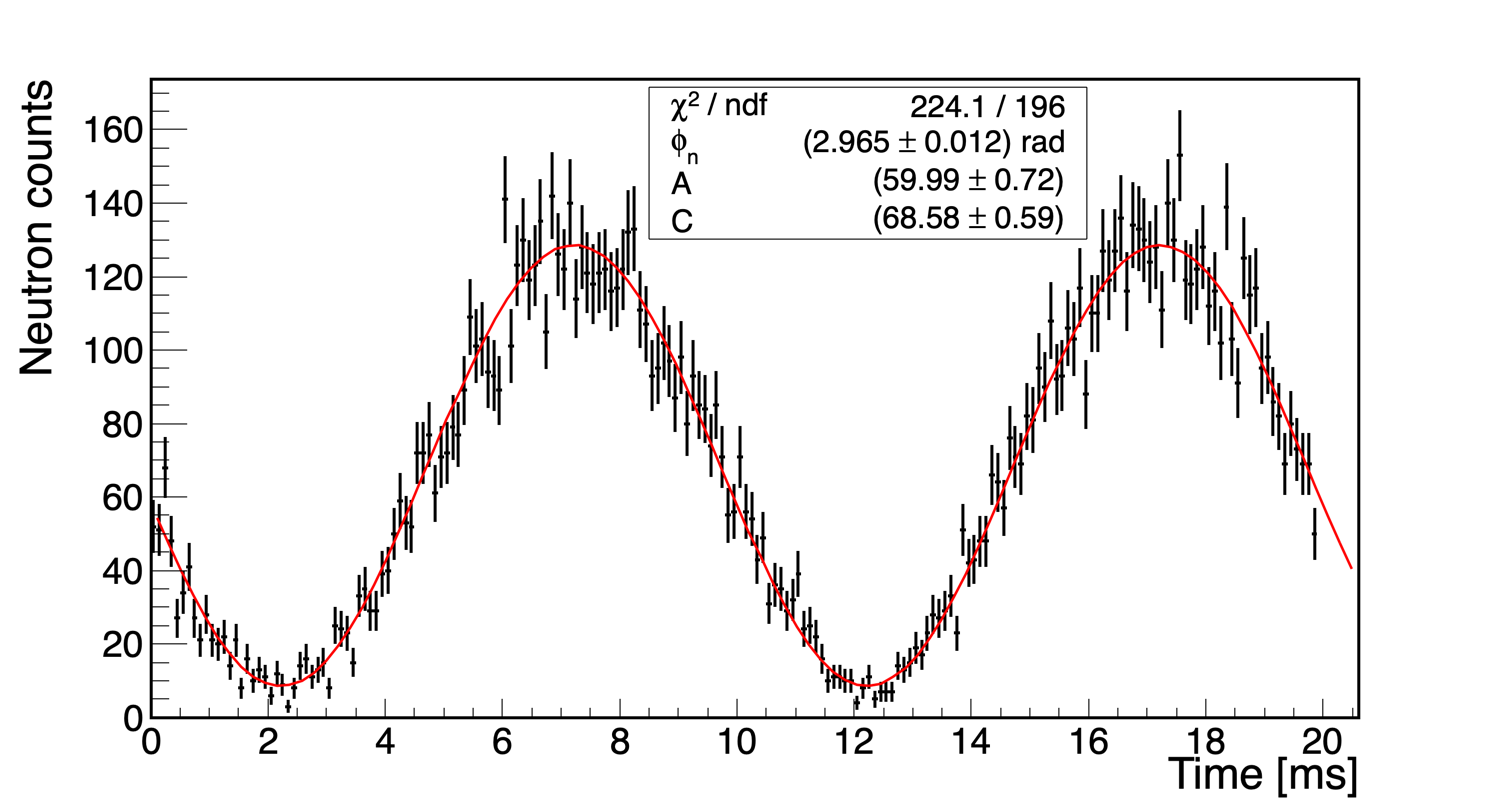}}
    
    \caption{FOSOF signals from the measurement with\newline $f_1=25050$~Hz, $f_2=24950$~Hz in (a) and $f_1=24950$~Hz, $f_2=25050$~Hz in (b). Neutron counts oscillate with the detuning frequency $\delta f=100$~Hz. The counts are grouped in bins of $100$~$\mu$s with two oscillation periods being recorded. 
    The solid lines correspond to sinusoidal fits to the data.
    }
    \label{fig:fosof_sig}
\end{figure}

\begin{figure} [h!]
    \centering
    \subfigure[]{\includegraphics[scale=0.225]{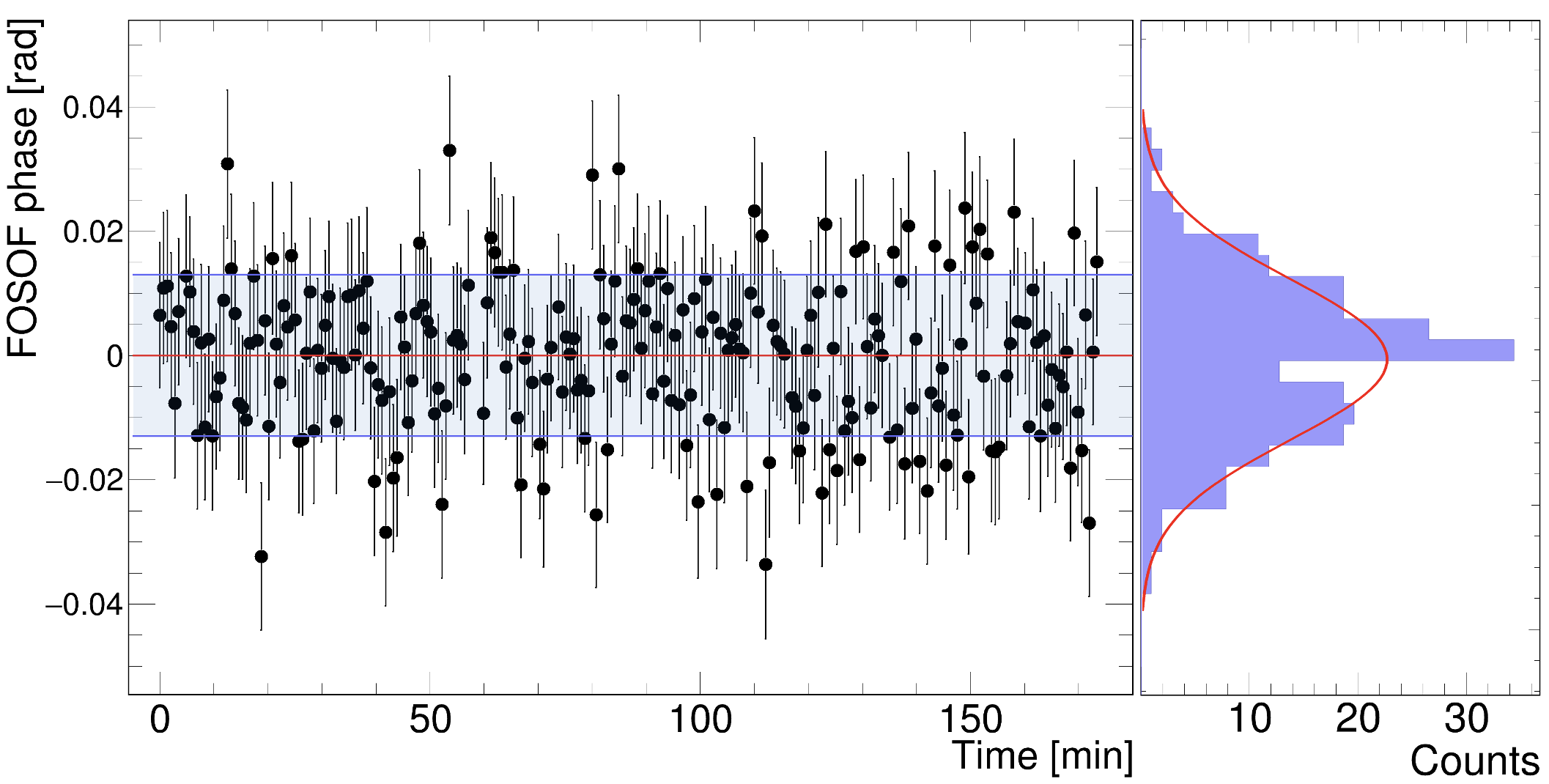}}
    \subfigure[]{\includegraphics[scale=0.17]{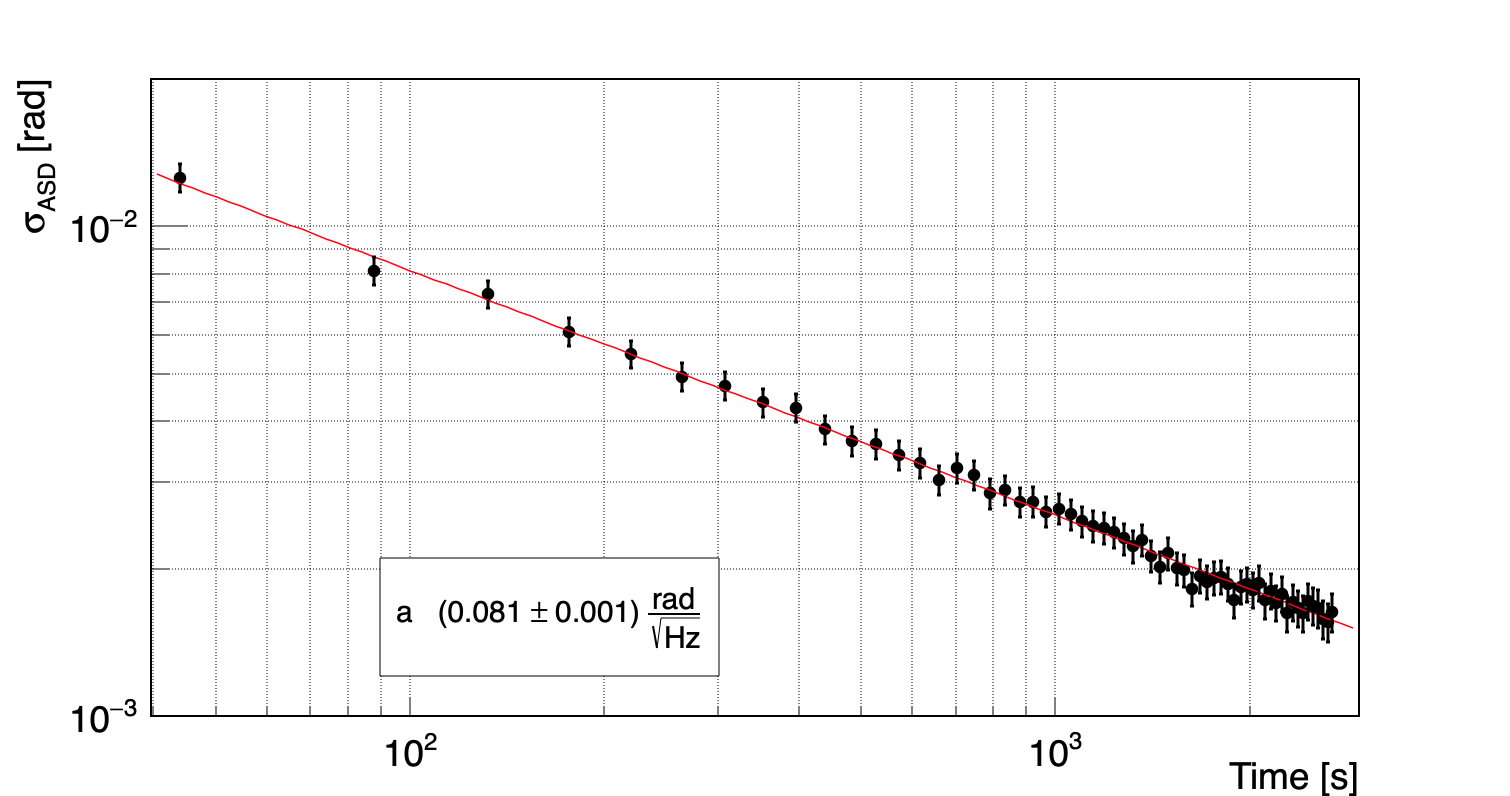}}
    \caption{Multiple consecutive FOSOF measurements with $f_1=24450$~Hz and $f_2=24550$~Hz have been performed over a period of about 3~hours. Plot (a) shows the FOSOF phase as a function of time with the related histogram. To enhance clarity, all phase values have been shifted vertically to obtain a mean value of zero. The horizontal blue shaded region around the mean value (red line) in the left plot describes $\pm 13$~mrad corresponding to the standard deviation of the Gaussian fit to the histogram. In (b) the related overlapping Allan standard deviation with a white-noise fit (red line) is shown.
    }
    \label{fig:phstab}
\end{figure}

In a FOSOF measurement, the neutrons are counted as a function of their detection time with respect to the trigger signal. 
Figure \ref{fig:fosof_sig} presents two exemplary FOSOF signals with neutron counts oscillating as a function of time with the detuning frequency $\delta f$. The FOSOF signals have been recorded during $2\times10^5$ neutron counts in the beam monitor detector corresponding to approximately 10~s of measurement time. 
Figure \ref{fig:fosof_sig}a shows the FOSOF signal acquired with $f_1 = 25050$~Hz and $f_2 = 24950$~Hz, while Fig.\ \ref{fig:fosof_sig}b shows the FOSOF signal acquired in the inverted frequency configuration with $f_1 = 24950$~Hz and $f_2 = 25050$~Hz. In both cases, the absolute value of $\delta f$ equals $100$~Hz. 
A sinusoidal fit to the data is performed using the function
\begin{equation}
 A \cdot \text{sin}(2\pi\cdot\delta f\cdot t-\phi_\text{n})+C,
\end{equation} 
where the parameters $A$, $C$, and $\phi_\text{n}$ are the amplitude, the offset, and the FOSOF phase, respectively. 
The presented FOSOF signals have an inferred modulation visibility of about 90\% with almost identical fitting parameters, however, they differ in their phases.
\newline To determine the sensitivity of the apparatus, a total number of 239 consecutive FOSOF measurements have been performed at constant spin-flip signal frequencies of $f_1 = 24450$~Hz and $f_2 = 24550$~Hz. 
Each measurement was approximately 10~s long and they were repeated every 45~s to allow for a readout of the data, leading to a total acquisition time of approximately 3 hours.
Figure~\ref{fig:phstab}a shows all corresponding FOSOF phases retrieved with a sinusoidal fit. 
From the presented histogram a phase sensitivity of $\pm 13$~mrad per measurement is obtained which agrees with the expected statistical uncertainty deduced from Monte Carlo simulations.
The phase shift $\Delta \phi_\text{n}$ and the magnetic field change $\Delta B$ are related via:
\begin{equation}
  \Delta\phi_\text{n} = - \gamma_\text{n} \cdot \Delta B \cdot T
\label{eq:larphi}
\end{equation}
where $\gamma_\text{n}$ is the neutron's gyromagnetic ratio 
\cite{gamma_n} and $T$ is the precession time of the neutrons.
This yields a corresponding magnetic field sensitivity of $\pm 105$~nT per measurement (assuming a precession time $T=650$~$\mu$s, compare Sec.\ \ref{par:fanalysis}).

Figure~\ref{fig:phstab}b depicts the derived Allan standard deviation (ASD) as a function of the integration time $\tau$  \cite{asd}. Here, the red line represents a white-noise fit to the data with the function $\frac{\text{a}}{\sqrt{\tau}}$, from which the parameter $a = (0.081 \pm 0.001)$~$\text{rad}/\sqrt{\text{Hz}}$ is determined.
Note, the graph exhibits no global minimum since the sensitivity of the apparatus is limited by neutron statistics.

\section{Frequency scan analysis}
\label{par:fanalysis}
\begin{figure}[]


\subfigure[]{\includegraphics[scale=0.17]{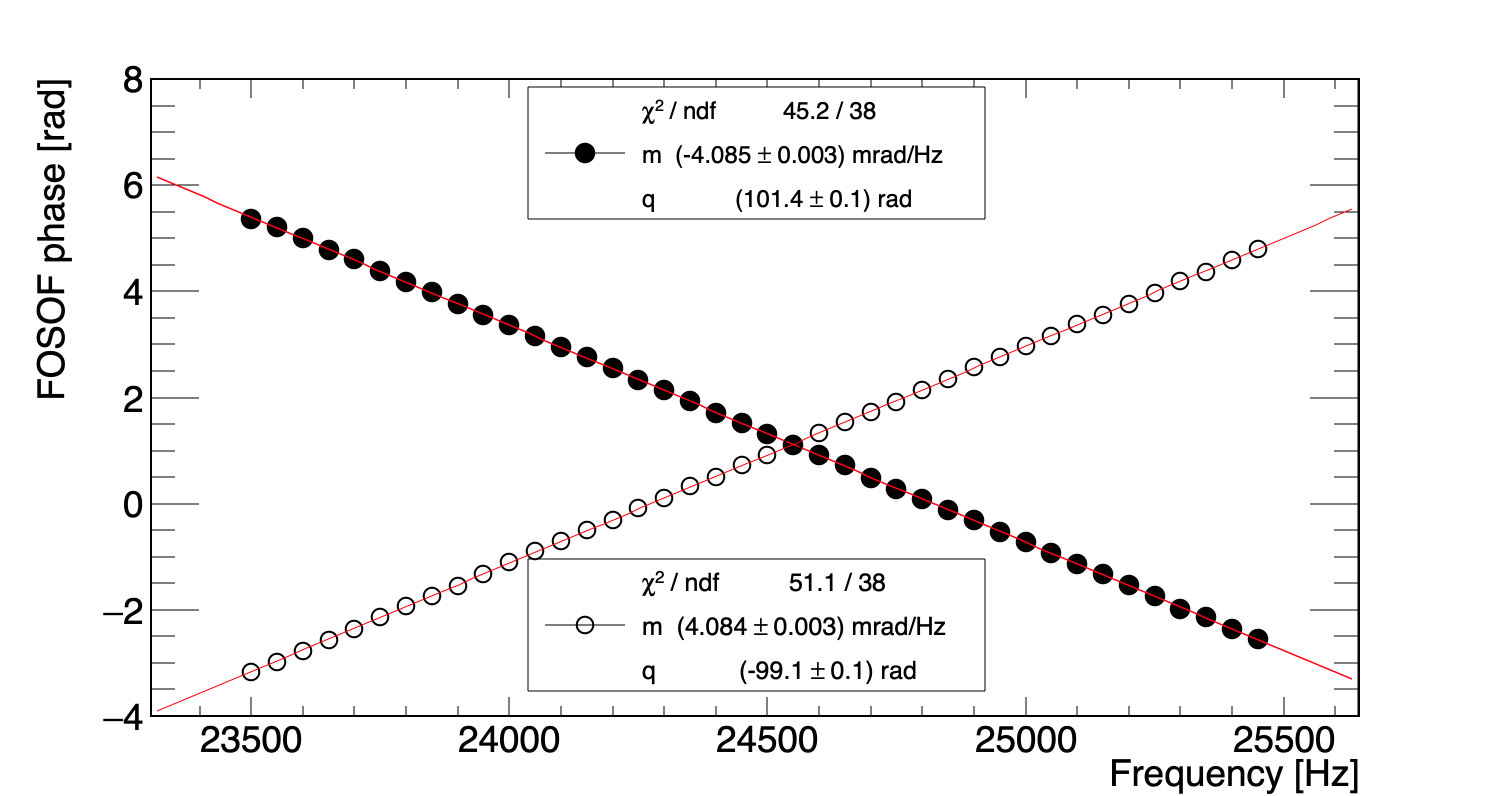}}
\subfigure[]{\includegraphics[scale=0.17]{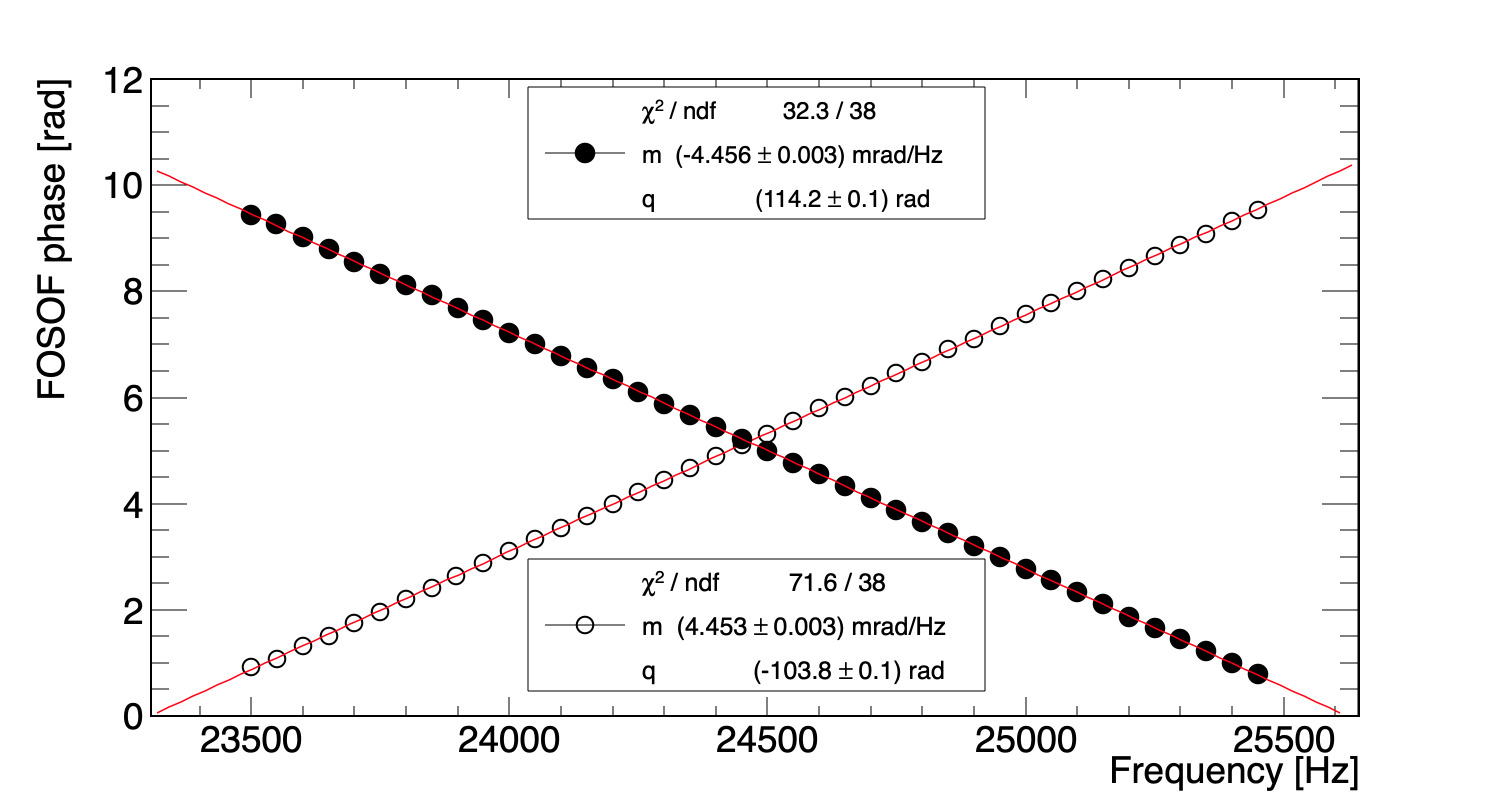}}
\subfigure[]{\includegraphics[scale=0.17]{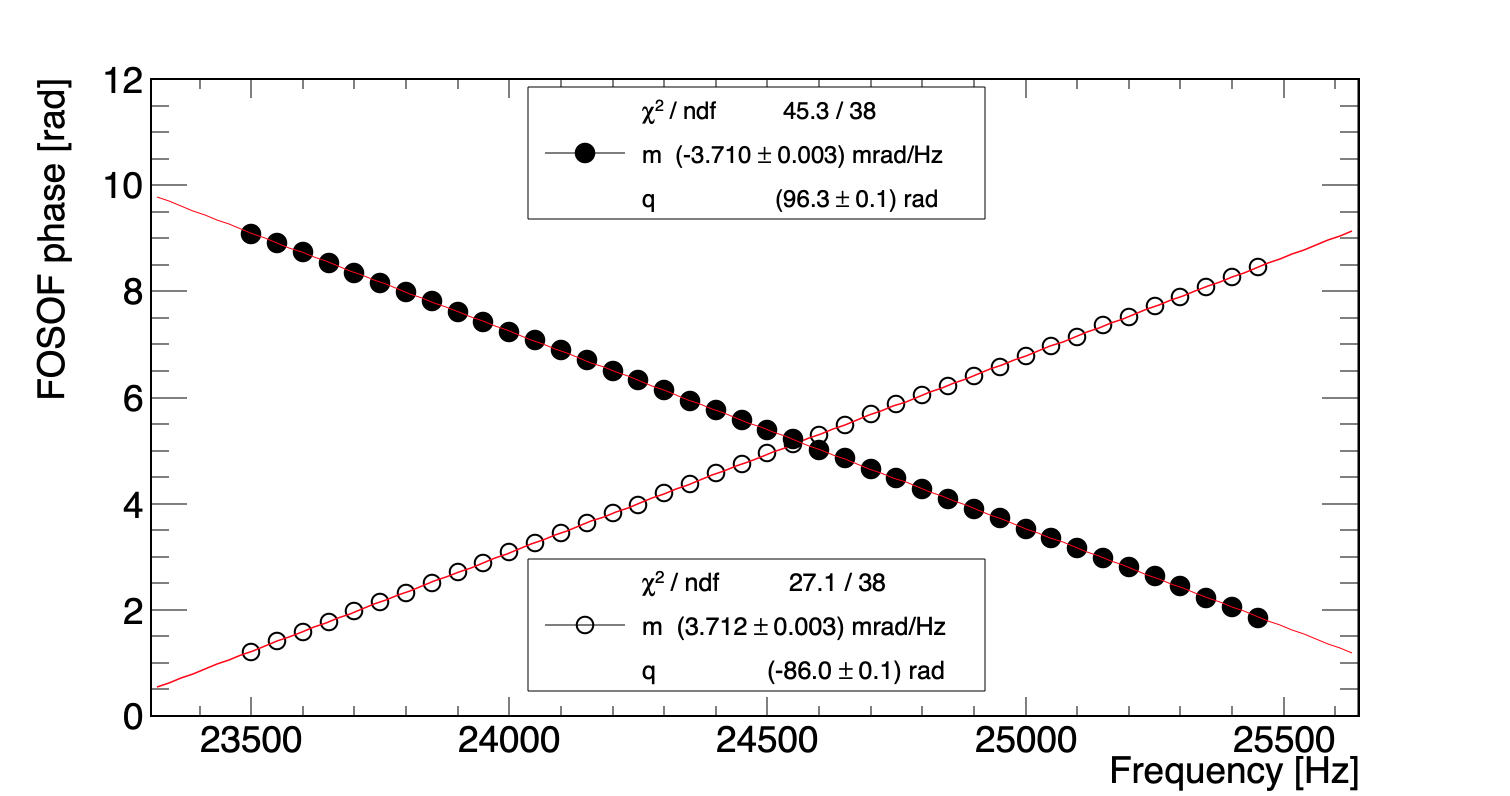}}

    \caption{FOSOF phase as a function of the scanned frequency $f$. They are obtained with three different precession region lengths: (a) shows the data with $L=500$~mm, (b) with\newline $L=550$~mm, and (c) with $L=450$~mm. The measurements were performed with $\delta f = 100$~Hz in the range between $23500$~Hz and $25450$~Hz. Empty circles and full circles 
    correspond to the measurement with $f_1 = f+\delta f/2$, $f_2 = f-\delta f/2$, and $f_1 = f-\delta f/2$, $f_2 = f+\delta f/2$, respectively.
    The solid lines represent the linear fits to the data.
    }
    \label{fig:cross_plots}
\end{figure}

As previously mentioned, the FOSOF measurement consists of two frequency scans: one with $f_1 = f+\delta f/2$, $f_2 = f-\delta f/2$, and the second one with inverted frequency settings $f_1 = f-\delta f/2$, $f_2 = f+\delta f/2$. 
From these two scans, it is possible to determine the resonance frequency by plotting the FOSOF phases as a function of $f$. Figure~\ref{fig:cross_plots}a shows the result of the frequency scans obtained with a precession region length $L=500$~mm and at a nominal magnetic field $B_{fluxgate}=~850$~$\mu$T. The resonance frequency is expected to be found close to $24500$~Hz. For this reason, the frequency has been scanned close to this value, in the range between $23500$~Hz and $25450$~Hz in steps of $50$~Hz. The empty circles represent the FOSOF phases measured with the normal frequency configuration, while the full circles represent the phases measured with the inverted configuration.
The vertical error bars are not visible, but are approximately $\pm 13$~mrad. 
A linear fit 
is used to evaluate the crossing point. As expected, the two slopes $m$ have the same value with opposite signs. The slope values depend on the precession region length $L$ and the neutron velocity. By taking into account the change of the Larmor precession frequency $\Delta f_\text{n} = - \frac{\gamma_\text{n}}{2\pi} \cdot \Delta B$ and employing Eq.\ (\ref{eq:larphi}) it follows:
\begin{equation}
    \frac{\Delta\phi_\text{n}}{\Delta f_\text{n}} = 2\pi\cdot T = 2\pi\cdot\frac{L}{v}
    \label{eq:slope}
\end{equation}
where the precession time $T$ is expressed as the ratio between $L$ and the neutron velocity $v$. By inverting Eq.\ (\ref{eq:slope}) it is possible to calculate the precession time from the measured slopes and compare it with the expected value. 
In this case, the slope of $(4.084\pm0.003)$~mrad/Hz results in a measured precession time of ($650.1\pm0.4$)~$\mu$s. 
To calculate the expected precession time for $L=500$~mm one must also account for the spin-flip coil length $l = 80$~mm.
Using the approximation: $L_{eff} \approx L + 0.27 \cdot l$, yields an effective precession region length  $L_{eff}=(522 \pm 1)$~mm \cite{Piegsa2008b}. The estimated uncertainty of the value originates from the mechanical positioning accuracy of the coils.
Additionally, assuming an estimated uncertainty of $1\%$ on the beam wavelength of $4.96$~$\text{\AA}$, which corresponds to a neutron velocity of $(798 \pm 8)$~m/s, the expected precession time is ($654\pm7$)~$\mu$s. 
Hence, the latter is compatible  with the measured value within one standard deviation.
The measurements have been repeated with a precession region length $L$ equal to $550$~mm and $450$~mm and are presented in Fig.\ \ref{fig:cross_plots}b and \ref{fig:cross_plots}c, respectively. The results are summarized in Table \ref{tab:slopes}.
\begin{figure} 
    \centering
    \includegraphics[scale=0.14]{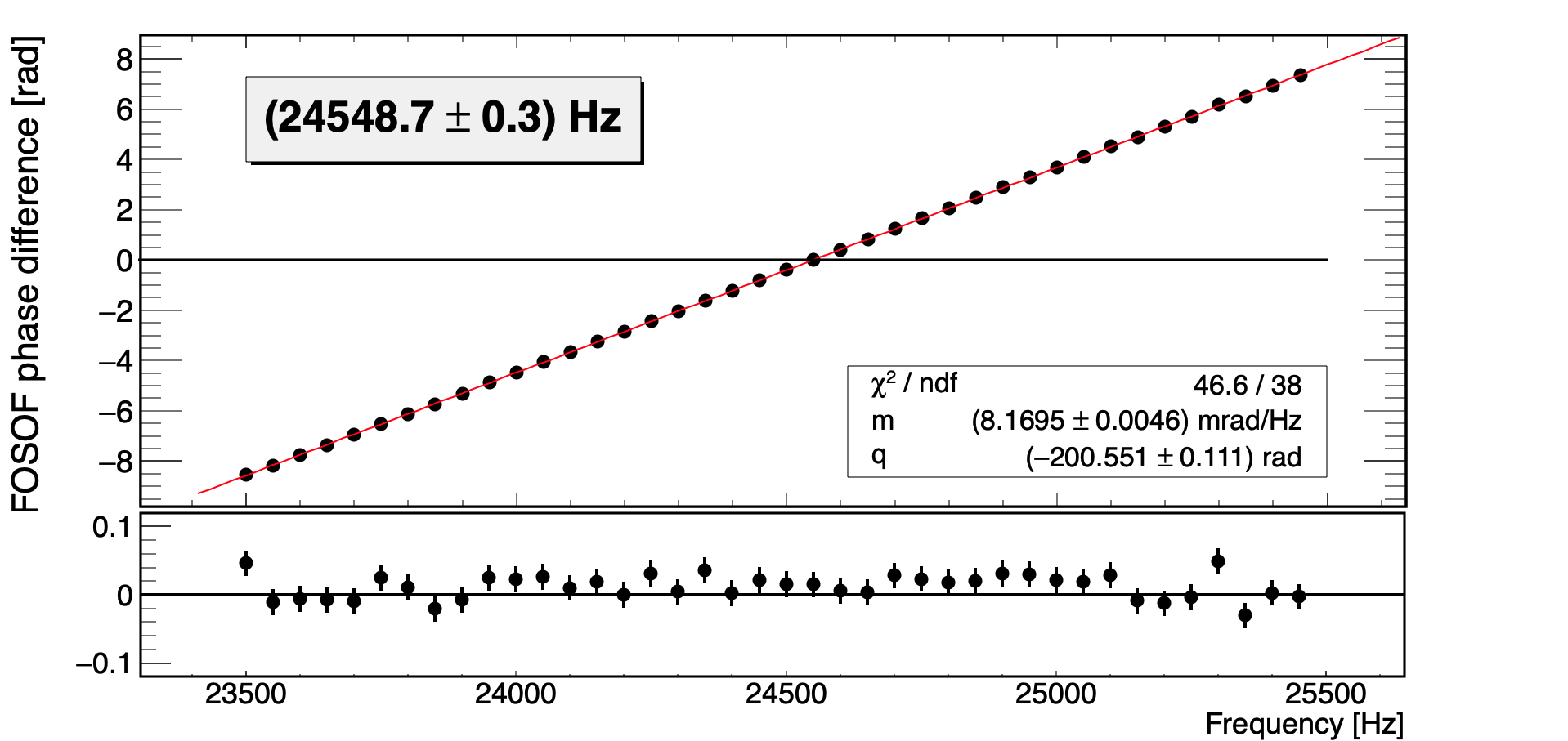}
    \caption{FOSOF phase difference for the case of $L=500$~mm employing the data shown in Fig.\ \ref{fig:cross_plots}a.
    The Larmor resonance frequency corresponds to the frequency where the FOSOF phase difference is equal to zero. The resulting value is ($24548.7\pm0.3$)~Hz. The residual plot from the linear fit is reported on the bottom. }
    \label{fig:resfreq50}
\end{figure}
\newline Once the scans with the two frequencies configurations
are performed, the Larmor resonance frequency is determined from the difference between the acquired phases. This is demonstrated in Fig.\ \ref{fig:resfreq50} where each data value is obtained from the phase difference between the FOSOF phases plotted in Fig.\ \ref{fig:cross_plots}a. The resonance frequency corresponds to the frequency of the point with the FOSOF phase difference equal to zero. A linear fit is performed and the fit parameters are used to calculate the resonance with its uncertainty. In this case, the resulting value is ($24548.7\pm0.3$)~Hz. 
The resonance frequency has been determined for several different detuning frequencies $\delta f$ and the results are reported in Table \ref{tab:df}. All measured values are compatible with each other within two standard deviations showing that the resonance frequency does not depend on $\delta f$.
    \begin{center}
\begin{table}[htp!]
    \centering
\resizebox{\columnwidth}{!}{    
    \begin{tabular}{ |c|c|c|c| }
     \hline
     $L$ [mm] & $L_{eff}$ [mm] & Exp. precession time [$\mu$s] & Meas. precession time [$\mu$s]\\
     \hline
     $500$ & $522 \pm 1$&$654\pm7$&$650.1\pm0.4$\\
     $550$ & $572 \pm 1$&$717\pm7$&$708.7\pm0.4$\\
     $450$ & $472 \pm 1$&$591\pm6$&$590.4\pm0.4$\\
     \hline
    \end{tabular}
    }
\caption{Expected and measured precession times for different precession region lengths $L$.}
\label{tab:slopes}
\end{table}
\end{center}

\begin{center}
\begin{table}[htp!]
    \centering
    \begin{tabular}{|c|c|c|}
     \hline
     $L$ [mm] & $\delta f$ [Hz] & Resonance frequency [Hz]\\
     \hline
     $500$ & $50$  & $24549.4\pm0.4$\\
     $500$ & $100$ & $24548.7\pm0.3$\\
     $500$ & $200$ & $24549.2\pm0.4$\\
     $500$ & $400$ & $24549.9\pm0.4$\\
           &       &                \\
     $450$ & $100$ & $24562.2\pm0.4$\\
     $550$ & $100$ & $24463.6\pm0.3$\\
     \hline
\end{tabular}
\caption{Larmor resonance frequencies measured for different detuning frequencies $\delta f$ and different precession region lengths $L$. For $L=450$ mm and $L=550$ mm, the average magnetic field experienced by the neutrons is slightly different from the one with $L=500$ mm resulting in a shift of the resonance frequencies.
}
\label{tab:df}
\end{table}
\end{center}

\section{Magnetic field scan}
\label{sec:mgscan}

\begin{figure}
    \centering
    \includegraphics[scale=0.17]{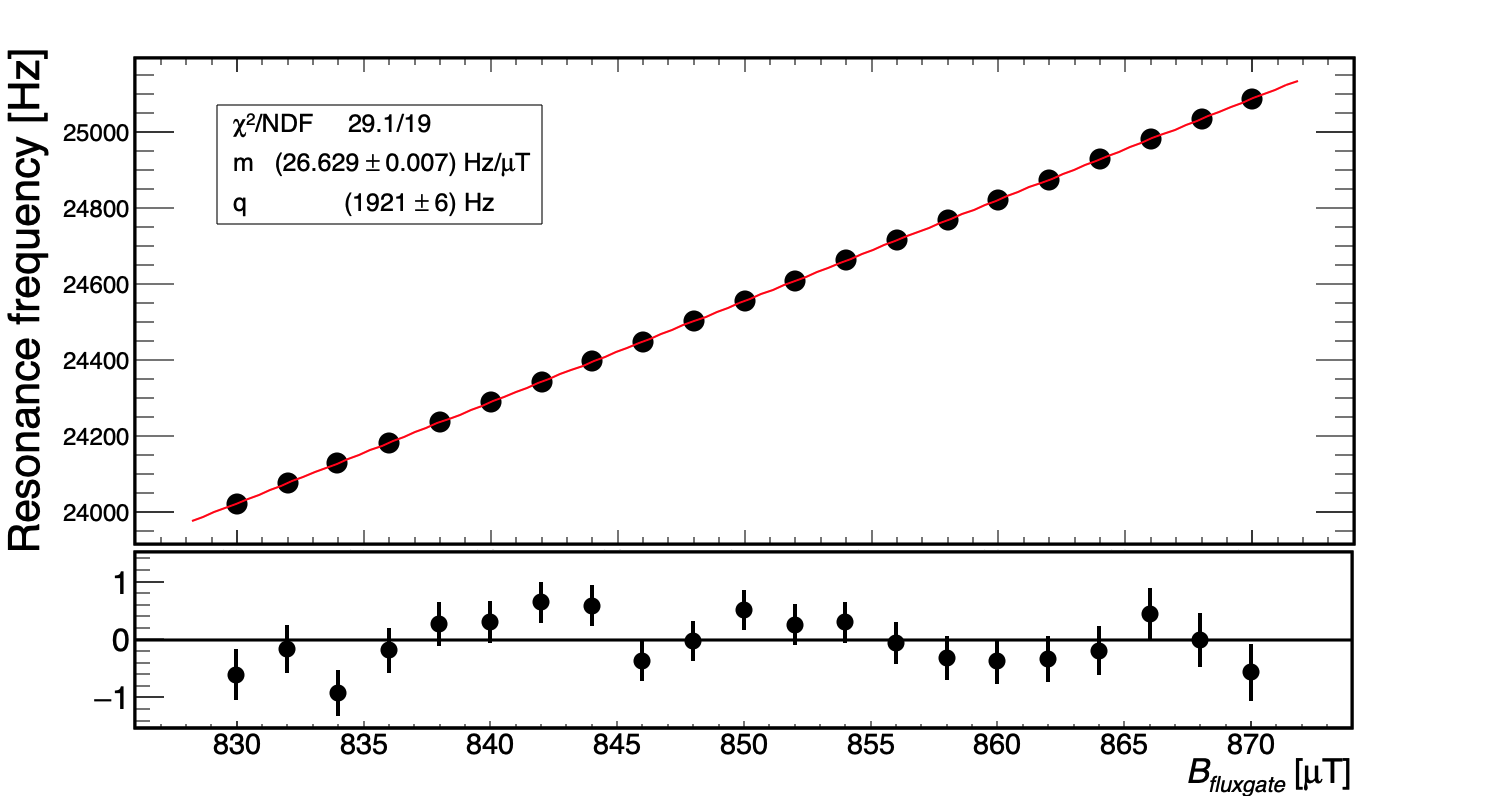}
    \caption{Larmor resonance frequency as a function of the nominal magnetic field at the fluxgate position. The measurement has been performed with the precession region length $L=500$~mm.  The residuals from the linear fit (solid line) with their error bars of order 0.3~Hz are presented on the bottom of the plot.
    }
    \label{fig:mgscan50}
\end{figure}

\begin{figure}
    \centering
    \includegraphics[scale=0.18]{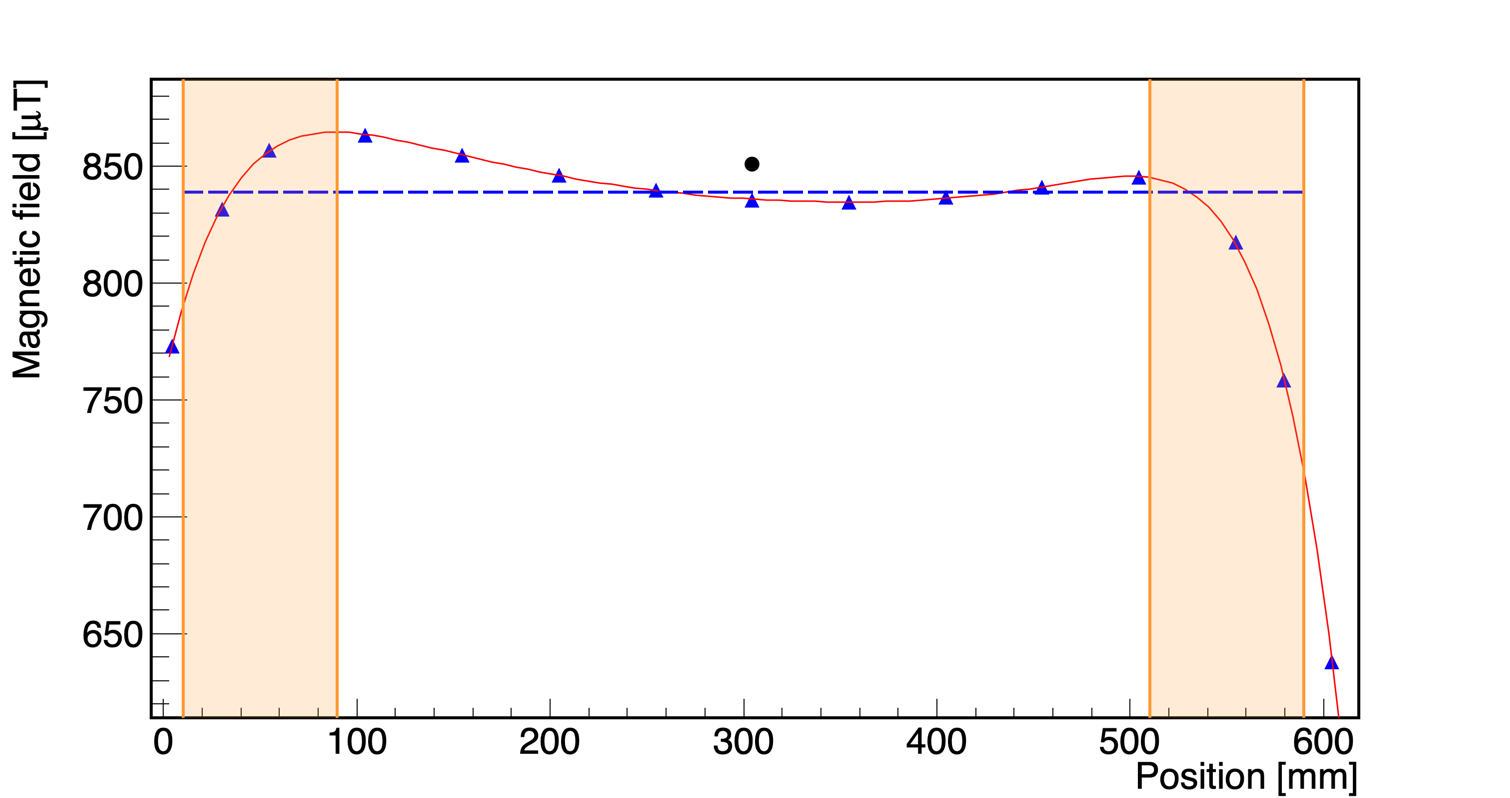}
    \caption{Measured magnetic field profile along the neutron beam direction (\textcolor{blue}{$\blacktriangle$}) and at the fluxgate position ($\CIRCLE$) at a nominal field of $B_{fluxgate} = 850$~$\mu\text{T}$.  
    The solid line represents an 8$^{\text{th}}$-order polynomial fit to the data. The shaded areas illustrate the positions of the spin-flip coils for a precession region length $L=500$~mm.
    The dashed horizontal line indicates the average magnetic field  $\overline{B}_0 = (839.1 \pm 0.3)$~$\mu\text{T}$ which the neutrons experience over the distance between 10~mm and 590~mm.
    }
    \label{fig:740mA}
\end{figure}

The Larmor resonance frequency depends linearly on the applied magnetic field in the precession region.
Figure \ref{fig:mgscan50} shows the result from a series of FOSOF measurements  for nominal magnetic fields $B_{fluxgate}$ in the range between $830$~$\mu$T and $870$~$\mu$T in steps of 2~$\mu$T. 
The data shows the anticipated linear behavior in accordance with the Larmor precession frequency formula: $f_n = - \frac{\gamma_n}{2\pi} \cdot B_0$, however, the fitted slope is equal to ($26.629\pm0.007$)~Hz/$\mu$T and thus not in agreement with the literature value for the neutron gyromagnetic moment $\lvert \frac{\gamma_\text{n}}{2\pi} \rvert = 29.1646931(69) $~$\text{Hz}/\mu\text{T}$ \cite{gamma_n}.
Moreover, the intercept of the fit is equal to ($1921\pm6$)~Hz instead of zero.
\newline
This discrepancy can be explained by the following considerations concerning the magnetic field.
First, the measurement and stabilization of the magnetic field is performed at one single point, namely, the position of the fluxgate, while the field experienced by the neutrons is averaged over the entire beam path. 
Second, the electromagnet providing the magnetic field exhibits a small gradient between the position of the fluxgate and the neutron path.
To investigate this, the magnetic field has been scanned along the beam direction using the same fluxgate. 
Figure \ref{fig:740mA} shows an example of such a magnetic field measurement along the beam path in comparison with the value at the fluxgate position. 
An interpolating fit with an $8^{\text{th}}$-order polynomial function has been applied to the data to extract the average magnetic field $\overline{B}_0$ seen by the neutrons between the outer edges of the spin-flip coils (for $L=500$~mm this corresponds to the distance between $10$~mm and $590$~mm). 
In the presented case a field $\overline{B}_0 = (839.1 \pm 0.3)$~$\mu\text{T}$ was determined. 
The stated uncertainty is dominated by the error on the fit parameters.
\begin{figure}
    \centering
    \includegraphics[scale=0.17]{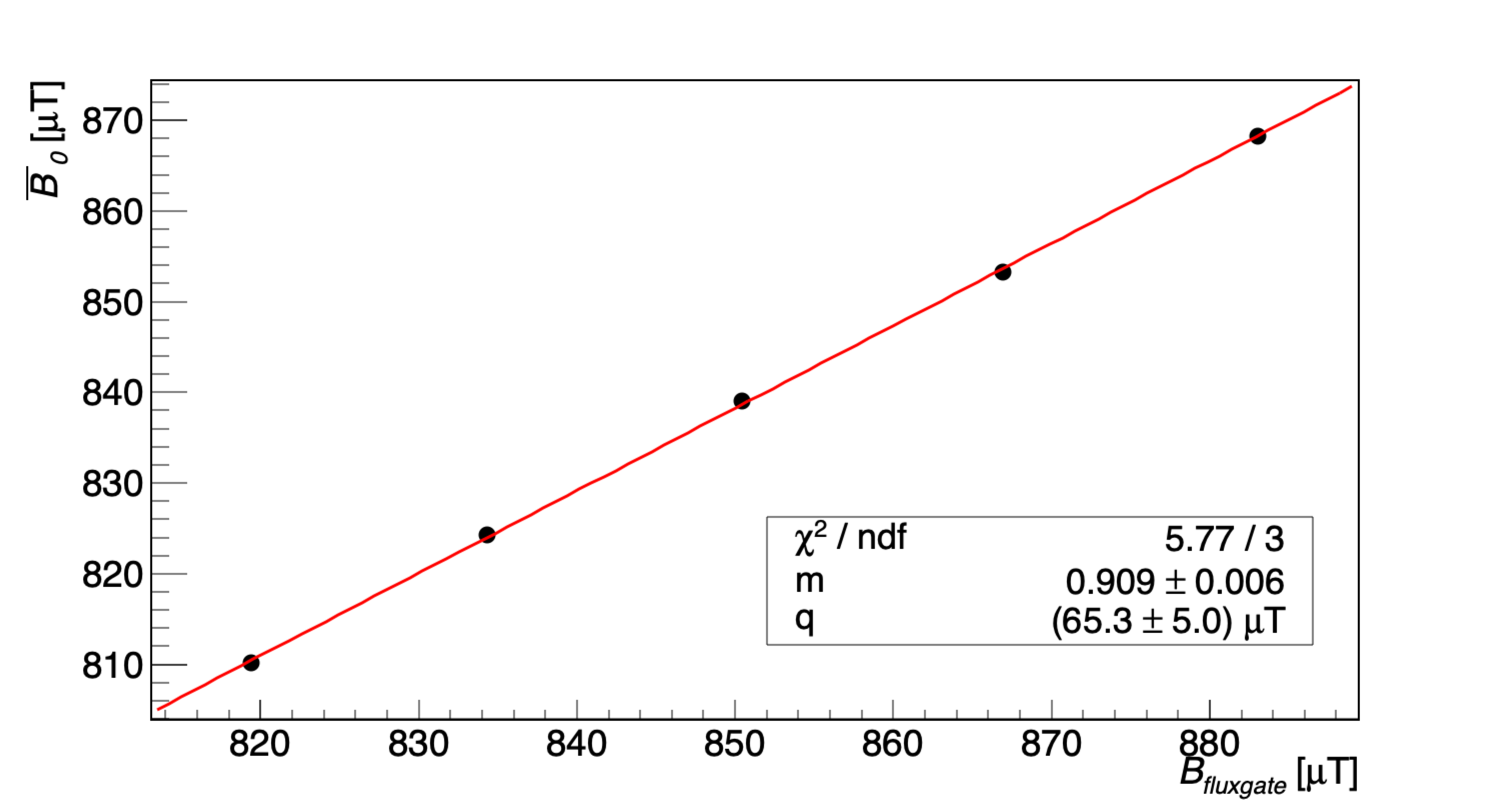}
    \caption{Average magnetic field $\overline{B}_0$ experienced by the neutrons as a function of the magnetic field at the fluxgate position $B_{fluxgate}$ with a precession region length $L=500$~mm.
    The solid line represents a linear fit to the data.
    }
    \label{fig:b0slope}
\end{figure}
\newline
The magnetic scanning has been repeated for different field settings and the results are presented in Fig.\ \ref{fig:b0slope}.
A linear fit yields two parameters, i.e.\ a field-correction factor of $(0.909 \pm 0.006)$ and an offset field value. 
The latter explains the aforementioned non-zero intercept: 
$\lvert \frac{\gamma_\text{n}}{2\pi} \rvert \cdot  (65.3 \pm 5.0)$~$\mu\text{T} = (1905 \pm 146)$~Hz.
Employing the field-correction factor results in a revised value for the slope in Fig.\ \ref{fig:mgscan50} of $(29.29 \pm 0.19)$~$\text{Hz}/\mu\text{T}$, which is now fully compatible with the literature value.
The increased uncertainty is caused by the approximate $0.7$\% relative error on the field-correction factor.




\section{Conclusions}

So far, the novel frequency-offset separated oscillatory fields technique has been applied in several high-precision measurements with atoms.
Here for the first time, the FOSOF technique has been adapted to a beam of polarized cold neutrons. Multiple characterization measurements have been performed at the spallation neutron source SINQ at the Paul Scherrer Institut.
FOSOF represents a useful alternative to the traditional Ramsey method and provides an addition to the portfolio in the realm of high-precision neutron spin precession techniques.
Potential applications concern experiments where an absolute resonance frequency is determined, e.g.\ the measurement of the gyromagnetic ratio and neutron gravity resonance spectroscopy.
Note, such high-precision experiments additionally require a meticulous consideration and mitigation of systematic errors and effects, e.g.\ magnetic field drifts, in order to achieve the desired accuracy. Moreover, absolute frequency measurements demand for the use of atomic clocks as stable frequency references. 




\section{Acknowledgement}
We gratefully acknowledge the excellent technical support by R. H\"anni, J. Christen, and L. Meier. 
The experiment has been performed at the Swiss Spallation Neutron Source SINQ at the Paul Scherrer Institut in Villigen, Switzerland.
This work was supported via the European Research Council under the ERC Grant Agreement no.\ 715031 (BEAM-EDM) and via the Swiss National Science Foundation under the grants no.\ PP00P2-163663 and 200021-181996.

\newpage

\nocite{*}
\bibliography{biblio}
\end{document}